\documentclass[prl,twocolumn,floatfix,amssymb,amsmath,secnumarabic,nofootinbib,
superscriptaddress,showpacs]{revtex4-1}

\usepackage{graphicx} 
\usepackage{dcolumn}  
\usepackage{bm}       
\usepackage{amsmath,amssymb}
\usepackage{natbib}
\usepackage{psfrag}
\usepackage{color}
\def\A{^{\sss A}}

\def\cc{^{\rm cc}}
\def\n{{\rm n}}

\def\sig#1#2{}

\def\bay{\begin{array}}
\def\eay{\end{array}}
\def\beit{\begin{itemize}}
\def\eit{\end{itemize}}

\def\bx{{\bf x}}

\def\d{\mathrm{d}}

\def\crap#1{The following is crap:\\ #1}
\def\crap#1{\bf !!!!Erroneous entry was removed here!!!!}
\def\bei{\begin{itemize}}
\def\eei{\end{itemize}}
\def\benu{\begin{enumerate}}
\def\enu{\end{enumerate}}

\def\tv{\tilde{v}}

\def\s{_{\rm s}}

\def\N{\tilde{N}}
\def\0{^{(0)}}

\def\1{^{(1)}}
\def\0{^{(0)}}
\def\del{\partial}

\newcommand{\brk}[3]{\langle #1\,|\ #2 \ |\,#3 \rangle}

\def\sig#1#2{\label{#1}\vskip -0.2cm{\small\em #1, {\bf #2}}\\}
\def\sig#1#2{\label{#1}}

\def\barr{\begin{array}}
\def\earr{\end{array}}

\newenvironment{fanc}
{\vskip 0.5 cm\rule{1ex}{1ex}\hspace{\stretch{1}}\noindent}
{\hspace{\stretch{1}}\rule{1ex}{1ex}\vskip 0.5cm}
\def\answ#1{}
\def\answ#1{~\\ \noindent{{\bf Answer:~}\rm\color{black} #1}}

\def\d{{\mathrm d}}

\def\N{{\cal N}}

\def\bea{\begin{eqnarray}}
\def\eea{\end{eqnarray}}
\def\ben{\begin{equation}}
\def\een{\end{equation}}
\def\benu{\begin{enumerate}}
\def\enu{\end{enumerate}}

\def\bei{\begin{itemize}}
\def\eei{\end{itemize}}
\def\beit{\begin{itemize}}
\def\eit{\end{itemize}}
\def\benu{\begin{enumerate}}
\def\enu{\end{enumerate}}

\def\n{n}

\def\sss{\scriptscriptstyle\rm}

\def\l{^\lambda}

\def\1var{(\bx_1...\bx\N)}

\def\half{\frac{1}{2}}

\def\br{{\bf r}}

\def\bx{{x}}

\def\s{_{\sss S}}
\def\xc{_{\sss XC}}

\def\N{_{\sss N}}
\def\H{_{\sss H}}

\def\TF{^{\rm TF}}

\def\ee{_{\rm ee}}

\def\sph_int{ {\int d^3 r}}

\def\tv{\tilde{v}}

\def\A{^{\sss A}}

\def\cc{^{\rm cc}}
\def\AD{^{\sss A, dir}}
\def\AV{^{\sss A, var}}

\begin{document}


\title{Electronic structure via potential functional approximations}

\author{Attila Cangi}
\affiliation{Department of Chemistry,
University of California, 1102 Natural Sciences 2,
Irvine, CA 92697-2025, USA}

\author{Donghyung Lee}
\affiliation{Department of Chemistry,
University of California, 1102 Natural Sciences 2,
Irvine, CA 92697-2025, USA}

\author{Peter Elliott}
\affiliation{Department of Physics and Astronomy, 
Hunter College and the City University of New York, 
695 Park Avenue, New York, NY 10065, USA}

\author{Kieron Burke}
\affiliation{Department of Chemistry,
University of California, 1102 Natural Sciences 2,
Irvine, CA 92697-2025, USA}

\author{E.K.U. Gross}
\affiliation{Max-Planck-Institut f\"ur Mikrostrukturphysik,
Weinberg 2, 06120 Halle (Saale), Germany}

\date{\today}

\begin{abstract} 
The universal functional of Hohenberg-Kohn is given as a
coupling-constant integral over the density as a functional of the {potential}.
Conditions are derived under which \emph{potential}-functional
approximations are variational.  Construction via this method and imposition
of these conditions are shown to greatly improve the accuracy of the non-interacting
kinetic energy needed for orbital-free Kohn-Sham calculations.
\end{abstract}


\maketitle


In the original form of density functional theory (DFT), 
suggested by Thomas\cite{T27} and Fermi\cite{F28} (TF)
and made formally exact by Hohenberg and Kohn\cite{HK64}, 
the energy of a many-body quantum system is minimized directly 
as a functional of the density.
Its modern incarnation uses the Kohn-Sham (KS) scheme\cite{KS65}, which employs
the orbitals of a fictitious non-interacting system.
This brilliant idea means only a small fraction
of the total energy need be approximated, and good approximations\cite{PBE96,Bb93}
have made DFT the popular tool it is today.  DFT\cite{FNM03} 
is now ubiquitous in many scientific fields, 
including both materials and chemistry.

Interest is rapidly reviving
in finding an {\em orbital}-free\cite{WC00}
approach to DFT. The major bottleneck in modern calculations is
the solution of the KS equations, which can be avoided with a
pure density functional for the kinetic energy
of non-interacting fermions, $T\s$.  The original TF approximation is
of exactly this type, but is far too inaccurate for modern applications.
Despite decades of effort\cite{DG90}, no generally applicable approximation for $T\s$
has been found, although material-dependent approximations\cite{HC10} have been
suggested, or approximations designed only for weakly-interacting
systems\cite{W08}.

However, Englert and Schwinger\cite{ES84} 
pointed out that the {\em potential} is a more
natural variable to use in deriving approximations to quantum systems.
In particular, semiclassical approximations begin with
the classical momentum, a local functional of the potential.
TF theory is often derived first in terms of the potential, which is
then eliminated in the final expressions, yielding an
explicit density functional.
Exact potential functional theory (PFT) satisfies a variational
principle with minimization 
over trial potentials\cite{YAW04,GP09}, yielding useful insight 
into the optimized effective potential method\cite{YAW04}.

In the present work we go beyond those results
by considering explicit potential functional approximations
to interacting and non-interacting systems of electrons;
such approximations are presently being developed
via a systematic asymptotic expansion 
in terms of the {potential}, which has already 
been found in simple cases\cite{ELCB08,CLEB10}.  The leading terms 
in a semiclassical expansion yield
local approximations to the energies, and the leading corrections 
greatly improve over the accuracy of local approximations in a systematic
and understandable way.   Corrections to TF are 
relatively simple functionals of the {potential},
but far more subtle as functionals of the density.  Such expansions are 
significantly more accurate and less problematic
for the density itself rather than for the kinetic energy density 
because the latter requires two spatial derivatives\cite{CLEB10}.

By minimizing over 
$N$-particle wavefunctions $\Psi$ that are 
antisymmetric, normalized, and have finite kinetic energy,
we obtain the ground-state (gs) energy 
\ben\label{E0psi}
E_v = \min_\Psi \left(\brk{\Psi}{\hat T + \hat V\ee + \hat V}{\Psi} \right)\,,
\een
as a functional of the potential, where
$\hat T$ is the kinetic energy operator,  
$\hat V\ee$ the electron-electron repulsion,
and $\hat V$ the one-body potential.
We define the potential functional\cite{YAW04}:
\ben
F[v] =
\brk{\Psi_v}{\hat T + \hat V\ee}{\Psi_v},\,
\een
with $\Psi_v$ denoting the gs wavefunction of potential $v(\br)$.
With $F[v]$, the exact relation for the gs energy is
\ben
E_v = F[v] + \int d^3r\ n[v](\br)\, v(\br)\,,
\een
in practice, requiring 
approximations to both $F[v]$ and $\n[v]$, 
\ben
E\AD_v = F\A[v] + \int d^3r\ \n\A[v](\br)\, v(\br)\,,
\label{Ead}
\een
where $A$ denotes an approximation as a functional of the potential.
We call this the direct approach.
  
We show that (i) the universal functional, $F[v]$, is determined entirely 
from knowledge of the density as a functional of the {\em potential},
such that only \emph{one} approximation is required, namely $n\A[v]$,
(ii) the variational principle imposes a condition relating energy and density
approximations, 
(iii) a simple condition guarantees satisfaction of the variational
principle, 
(iv) with an orbital-free approximation to the non-interacting density as a functional 
of the {\em potential}, the kinetic energy is automatically determined, 
i.e., there is no need for a separate approximation,
and (v) satisfaction of the variational principle improves accuracy of 
approximations.

We \emph{deduce} 
an approximation to $F$ from any $\n\A[v](\br)$
in the following way.
Introduce a coupling constant in the one-body potential:
\ben\label{potcoup}
v\l(\br)  = (1-\lambda)\, v_0(\br) + \lambda\, v(\br)\,,
\een
where $v_0(\br)$ is some reference potential (possibly $0$). 
In the context of TF theory
this coupling was used in Ref.~\cite{PHH88}.
Then, using the Hellmann-Feynman theorem,
\ben
E_v = E_0 + \int_0^1 d\lambda\, \int d^3r\
      \n[v\l](\br)\, \Delta v(\br)\,,
\een
where $\Delta v(\br) = v(\br) - v_0(\br)$.
Defining  $\bar{n}[v](\br) = \int_0^1 d\lambda\ n[v^\lambda](\br)$
and choosing $v_0(\br)=0$, we obtain
\ben\label{cc}
F[v] = \int d^3r \left\{ \bar{n}[v](\br)-\n[v](\br) \right\}\; v(\br) \ .
\een
This formula establishes that the universal functional is determined 
solely by the knowledge of the density as functional of the {\em potential}.
Moreover, insertion of $\n\A(\br)$ on the right {\em defines} an associated 
approximate $F\cc[n\A[v]]$, where $cc$ denotes coupling constant.

On the other hand,
much of the accuracy of DFT calculations derives from the variational principle.
In PFT, this yields
\ben
E\AV_v = \min_{\tv} \left( F\A[\tv] + \int d^3r\ \n\A[\tv](\br)\, v(\br)\right),
\label{Ev}
\een
with a possibly different value from Eq.~(\ref{Ead})
for a given pair of approximations.
Experience suggests 
use of the variational principle improves results.
The Euler equation for the minimum
is
\ben\label{euler}
\frac{\delta F\A[v]}{\delta v(\br)}
= -\int d^3r'\ v(\br')\, \chi\A[v](\br',\br),
\een
where $\chi\A[v](\br,\br')=
\delta n\A[v](\br)/\delta v(\br')$ denotes the density-density
response function.
If a pair of approximations
$\left\{F\A, \n\A\right\}$ satisfies Eq.~(\ref{euler}) at $v(\br)$,
then Eqs.~(\ref{Ead}) and (\ref{Ev}) yield identical results,
but this is {\em not} guaranteed
{\em a priori} in approximate PFT.

We next ask: Does a given $F\cc[\n\A[v]]$ satisfy
Eq.~(\ref{euler})?  Taking the functional derivative of Eq.~(\ref{cc}) yields
Eq.~(\ref{euler}) if, and only if,
\bea
\n\A[v](\br&)=&\int_0^1 d\lambda\int d^3r'\ \Big\{ \frac{\delta \n\A[v\l[v]](\br')}{\delta v(\br)}\,
    \frac{d v\l[v](\br')}{d\lambda}\nonumber\\
&+& \n\A[v\l[v]](\br')\,
    \frac{d}{d\lambda}\frac{\delta v\l[v](\br')}{\delta v(\br)} \Big\}.
\eea
This condition is true in turn, if and only if,
\ben\label{symrel}
\chi\A[v](\br,\br') = \chi\A[v](\br',\br)\ .
\een
The exact response function satisfies this relation, 
but an approximate functional $n\s\A[v]$ might not.
This condition guarantees conservation of particle number
under small changes in the potential, an elementary version
of a conserving approximation\cite{BK61}.

A simple example illustrating these results is TF theory, considered
as a potential functional.  Then,
\ben\label{nTF}
\n\TF[v](\br) = \frac{1}{3\pi^2}\, \left[2(\mu-v\s(\br))\right]^{3/2}\,,
\een
where $v\s(\br)=v(\br)+v\H(\br)$, and the latter is the Hartree
potential, determined self-consistently from 
\ben
\nabla^2 v\H(\br) = -4\pi\, n\TF(\br)\,,
\een
while $\mu$ is the chemical potential, determined by the normalization
requirement that $\int d^3r\ \n(\br)=N$.  Taking functional
derivatives with fixed $N$,
the usual TF energy expression\cite{CB11}
\ben
F\TF[v] = T\s\TF[v] + \int d^3r\, n\TF[v](\br)\ v(\br) + U[n\TF[v]]
\een
satisfies the Euler condition when combined with Eq.~(\ref{nTF}), 
and $F\cc[\n\TF[v]]=F\TF[v]$, where
$T\s\TF[v]=\pi^2\int d^3r\ n\TF[v](\br)^3/6$ denotes the 
TF kinetic energy and
$U[n\TF[v]] = \int d^3r\ n\TF[v](\br)\ v\H[n\TF[v]](\br)/2$
the Hartree energy.

In practice, the usefulness of these results for interacting electrons
might be limited, as they require an approximation to
the \emph{interacting} density as a functional of the one-body potential that
is sufficiently accurate to be competitive with standard KS-DFT 
calculations, i.e., beyond the accuracy of 
TF theory.  Of much more practical use is their
application to the non-interacting electrons of the KS scheme, 
which sit in the effective KS potential, which includes both a Hartree 
and (some approximate) exchange-correlation (XC) contribution:
\ben\label{vspft}
v\s(\br) = v(\br) + \int\d^3r'\ \frac{\n\A\s[v\s](\br')}{|\br-\br'|}
           + v\xc\A[\n\A\s[v\s]](\br)\,.
\een
For a given approximation to $E\xc$, which determines $v\A\xc$,
this equation can be easily solved by standard iteration techniques, 
{\em bypassing} the need to solve the KS equations.
A given $n\s\A[v\s]$ removes the need
for solving any differential equation in each iteration.

However, once self-consistency is achieved, we need to extract the total
energy of the interacting electronic system, for which we need the kinetic
energy of the KS electrons.  All our derivations apply equally to
the non-interacting problem, so we deduce:
\ben\label{Ts}
T\s[v] = \int d^3r \left\{ \bar{n}\s[v](\br)-\n\s[v](\br) \right\}\; v(\br) \,,
\een
which is the analog of Eq.~(\ref{cc}) for a system of non-interacting
electrons in the external potential $v(\br)$
(which is called $v_s(r)$, 
when it is the KS potential of some interacting system).
This {\em defines} a kinetic energy approximation determined solely
by the density approximation:
\ben\label{Tscc}
T\s\cc[\n\s\A[v]]
= \int d^3r \left\{ \bar{n}\s\A[v](\br) - \n\s\A[v](\br) \right\}\, v(\br)\ .
\een
This is our main result for the non-interacting case.
It eliminates the need for constructing
approximations to the non-interacting kinetic energy $T\s$.

To illustrate the power of these results, we 
consider a simple example,  
a system of non-interacting, spinless fermions
in a one-dimensional box.  We choose $v_0(x)$ to be 
$0$ inside a box ($0 < x < L$),  and $\infty$ outside.
Then $\Delta v(x) = v(x)$ is some potential inside the box.
For this case Eq.~(\ref{Tscc}) (with a nonzero $v_0$) reduces to 
\ben\label{Ts1dbox}
T\s\cc[\n\s\A[v]] = E_0(N) +\! 
          \int_0^L\!\!\!\!\! dx 
	  \left\{ \bar{\n}\s\A[v](x)  -n\s\A[v](x) \right\} v(x)\,,
\een
where $E_0(N)=\pi^2\,\left( N^3 + 3/2\, N^2 + N/2 \right)/6$
denotes the total non-interacting energy of $N$ spinless fermions 
in an infinite square well. 

\begin{figure}[htp]
\begin{center}
\includegraphics[angle=0,height=6cm]{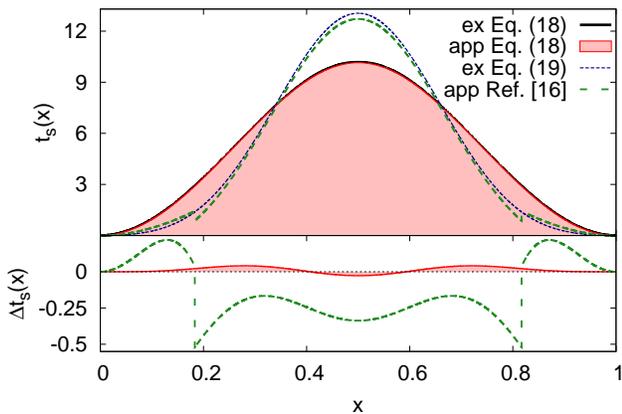}
\end{center}
\caption{Exact (ex) and approximate (app) kinetic energy densities (above)  
of Eq.~(\ref{Ts1dbox}) (black, red) and of
Eq.~(\ref{ts1d}) (blue, green) with the approximation
in Ref.~\cite{CLEB10},
and their absolute errors (below) for one particle 
in $v(x) = -5\, \sin^2 (\pi x)$, $0~<~x~<~1$. (color online)}
\label{f:ts}
\end{figure}
In Fig. \ref{f:ts} we plot two distinct kinetic energy densities,
along with approximations to them, and the corresponding errors, for
$v(x)=-5 \sin^2 (\pi x)$ in a box of unit length.
The blue curve is the exact kinetic energy density obtained from
a traditional definition,
\ben\label{ts1d}
t\s(x) = -\half \sum_{j=1}^N \phi_j^*(x)\frac{\del^2}{\del x^2} \phi_j(x)\,, 
\een
while the green curve is the approximation derived at great length in
Ref.~\cite{CLEB10}.  The small discontinuity at about $x=0.2$ and $0.8$
is where the approximation switches from a form that is asymptotically
correct in the interior to one that is asymptotically correct near the
walls.  The error is shown in the bottom panel.  Note that the 
approximation for $t\s(x)$ of Ref.~\cite{CLEB10} is already a considerable
improvement over that used in Ref.~\cite{ELCB08}.
The red and black curves result from Eq.~(\ref{Ts1dbox}).
The black is exact, while the red uses the approximation for the density
in Ref.~\cite{ELCB08}.  Their difference is plotted in the bottom panel,
and is both locally and globally far smaller, and
required no separate approximation for the kinetic energy density.

The approximations of Refs.~\cite{ELCB08} and \cite{CLEB10} were
designed to be asymptotically exact as $N\to\infty$, both for the
density and the kinetic energy.  
In Table~\ref{t:Ts_sinlD5} we show errors compared to the exact result 
of $T\s\A$ from TF theory, the WKB approximation,
Ref.~\cite{ELCB08}, its improvement in Ref.~\cite{CLEB10}, 
and $T\s\cc$ in Eq.~(\ref{Ts1dbox}) with $\n\s\A[v]$ of Ref.~\cite{ELCB08}.
\begin{table}[htb]
\caption{Total non-interacting kinetic energy of N particles and
its absolute error in TF, WKB, $T\s\A$ of Ref.~\cite{ELCB08} and 
\cite{CLEB10}, and $T\s\cc$ of Eq.~(\ref{Ts1dbox}) for the potential 
of Fig.~\ref{f:ts}.}
\label{t:Ts_sinlD5}
\begin{ruledtabular}
\begin{tabular}{c|r|rcccc}
 & & \multicolumn{5}{c}{$|T\s-T\s\A|$}\\
\hline
$N$ & $T\s$ & TF & WKB & Ref.~\cite{ELCB08} & Ref.~\cite{CLEB10} & $T\s\cc$\\
\hline
 1 &   4.97 &  3.16 & 1.42 & 0.47 & 0.12 &  $1.2 \cdot 10^{-2}$\\
 2 &  24.73 & 11.50 & 1.47 & 0.43 & 0.08 &  $2.0 \cdot 10^{-3}$\\
 4 & 148.08 & 42.76 & 1.48 & 0.50 & 0.04 &  $4.0 \cdot 10^{-4}$\\
\end{tabular}
\end{ruledtabular}
\end{table}
Even for $N=1$, $T\s\cc$ is about 
two orders of magnitude more accurate than WKB, 
and significantly more accurate than the direct approximations 
of Refs.~\cite{ELCB08} and \cite{CLEB10}. 
As $N\to\infty$, $T\s\cc$ converges most rapidly.
\begin{figure}[htb]
\begin{center}
\includegraphics[angle=0,height=6cm]{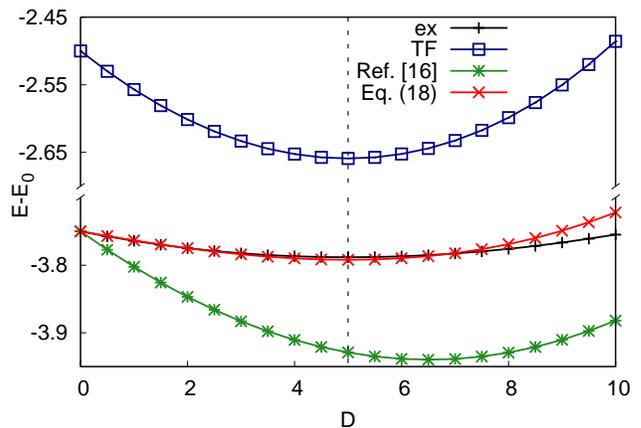}
\end{center}
\caption
{Difference of total exact (ex) energy $E$ and $E_0$ (black) 
compared to the TF approximation (blue), $T\s\A$ 
of Ref.~\cite{CLEB10} (green) and $T\s\cc$ of Eq.~(\ref{Ts1dbox}) (red) 
for $N=1$ and trial potentials $\Delta \tv(x)=-D\sin^2(\pi x)$. (color online)}
\label{f:dDEE0_N1}
\end{figure}

We finally test the symmetry condition of Eq.~(\ref{symrel}). 
To do this, we perform a variational PFT calculation, implementing
Eq.~(\ref{Ts1dbox}).  We take a given external potential ($-5 \sin^2 (\pi x)$),
calculate exact gs wave functions with different potentials,
and find their energy.  Fig. \ref{f:dDEE0_N1} shows the
results when the well depth $D$ is varied.  The
exact result is a black curve, whose minimum occurs at $D=5$.
The blue curve is the result of TF theory, which satisfies the condition,
but is not very accurate.  The green curve is the approximation of Ref.~\cite{CLEB10},
which, while more accurate, does not minimize at the true potential. 
This demonstrates that the pair of approximations $T\s\A[v]$ and $\n\s\A[v]$
given there do {\em not} yield the same answer variationally and directly.
Note also that the direct evaluation (green curve at $D=5$)
is more accurate, when compared to the exact result (black curve at $D=5$)
than application of the variational principle 
(green curve at its mininum $D\approx 6.5$).  
This is because these approximations
were derived semiclassically, and have uncontrolled errors.
On the other hand, our coupling-constant approximation of Eq.~(\ref{Ts1dbox}) 
(red curve) is both far more accurate and minimizes at the true potential.  
\begin{figure}[htb]
\begin{center}
\includegraphics[angle=0,height=6cm]{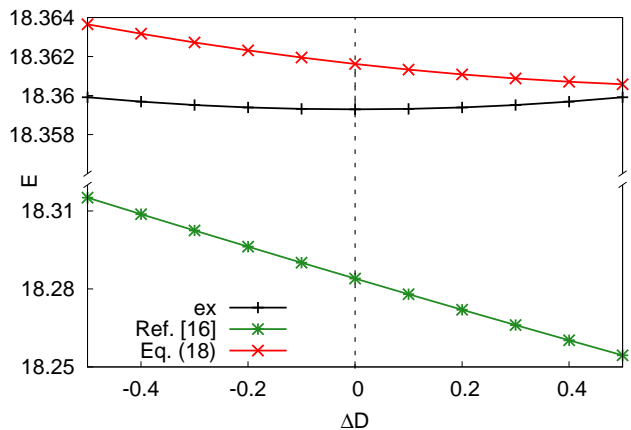}
\end{center}
\caption
{Exact (ex) total energy (black) compared to 
the direct approximation of Ref.~\cite{CLEB10} (green)
and $T\s\cc$ of Eq.~(\ref{Ts1dbox}) (red) for $N=2$ and 
trial potentials $\tv(x)=-5\sin^2(\pi x) - \Delta D\sin^2(2\pi x)$ 
with the external potential at $\Delta D=0$. (color online)}
\label{f:DEsd_N2}
\end{figure}

However, we further tested the coupling-constant approximation 
by adding the perturbation $-\Delta D\sin^2 (2\pi x)$ with
varying depth $\Delta D$ to the given external potential ($-5 \sin^2 (\pi x)$). 
For $N=2$, as shown in Fig.~\ref{f:DEsd_N2},
the minimum is no longer at $\Delta D=0$, despite the high accuracy.
This demonstrates that Eq.~(\ref{symrel}) is {\em not} satisfied for all
possible variations of the potential around the true one.   
However,
the breakdown appears small and does not much diminish accuracy.
This breakdown is due to the small normalization error
in the semiclassical density\cite{CLEB10}.  
If the error changes with the potential, 
the corresponding $\chi^A$ cannot be symmetric.
We have also calculated the response function 
for that approximation and found non-symmetrical terms\cite{CB11}.
This shows the utility of our analysis:  Accuracy is likely to be
further improved if the result can be easily modified to {\em enforce}
Eq.~(\ref{symrel}).

An alternative to the coupling-constant method used here is 
the virial theorem, which yields the kinetic energy from the potential
and density alone\cite{MY59}.   We recommend that version which
has an origin-independent kinetic energy density given in Ref.~\cite{SLBB03},
satisfying 
\ben
\nabla^2 t\s(\br) = -\frac{d}{2} \nabla\{ n(\br)\,\nabla v\s(\br) \}\,,
\een
where $d$ denotes the dimension of space.
While either the virial or the coupling-constant formulation can
be applied to realistic systems, we use the coupling constant here because
our illustrations involve box-boundary conditions, which create complications
for the virial theorem\cite{CP51}.   

The coupling-constant
formulation can be applied to realistic systems with potentials that vanish
at large distances, using $v_0=0$ for a reference. Then, to 
keep the particle number fixed, employ the device of putting
the system in a large box whose size is taken to $\infty$ at the end of
the calculation.  
Either expression has a great advantage over traditional 
density-functional approximations,
such as generalized gradient approximations.
For such approximations, 
there is always an ambiguity in the energy densities; a term that
integrates to zero over the entire space can always be added\cite{PSTS08}.
However our energies use an approximation to the density, 
which is uniquely determined for all $\br$, and so can be used to
identify the relative contribution to the energy from different regions\cite{CB11}.

Which variation (coupling constant or virial) is most useful in practice
awaits general-purpose approximations for the density
as a functional of the potential for an arbitrary three-dimensional
case.  But at least it no longer awaits the corresponding kinetic
energy approximations.

We gratefully acknowledge funding from NSF under 
grant number CHE-0809859.

\bibliography{Master}
\label{page:end}
\end{document}